\documentclass[doublecol,linenumbers]{epl2} 
\usepackage{amssymb}   % 

\usepackage{graphicx}

\usepackage{epstopdf}

\usepackage{amssymb}   

\usepackage{amsmath}

\title{Ultra long lived quasinormal modes of neutron stars in massive scalar-tensor gravity
}
\shorttitle{Ultra long lived quasinormal modes of neutron stars in massive scalar-tensor gravity
}

\author{Jose Luis Bl\'azquez-Salcedo\inst{1} \and Fech Scen Khoo\inst{2} \and Jutta Kunz\inst{1}}
\author{Jose Luis Bl\'azquez-Salcedo, Fech Scen Khoo, Jutta Kunz}

\institute{                    
	\inst{1} Institut f\"ur  Physik, Universit\"at Oldenburg  Postfach 2503,
	D-26111 Oldenburg, Germany\\
	\inst{2} Spoken Language Systems Group, Saarland University,
	Saarland Informatics Campus, 66123 Saarbr\"ucken, Germany
}


\abstract{
	The spectrum of frequencies and characteristic times 
	that compose the ringdown phase of gravitational waves 
	emitted by neutron stars carries information about the matter content 
	(the equation of state) and the underlying theory of gravity. 
	Typically, modified theories of gravity introduce 
	additional degrees of freedom/fields, such as scalars,
	which result in new families of modes composing the ringdown spectrum. 
	Simple but physically promising candidates are scalar-tensor theories,
	which effectively introduce an additional massive scalar field (i.e. an ultra-light boson)
	that couples non-minimally to gravity, resulting in scalarized neutron stars. 
	Here we present the first calculation of the full ringdown spectrum in
	such theories. We show
	that the ringdown spectrum of neutron stars with ultra-light bosons 
	is much richer and fundamentally different from the spectrum in 
	general relativity and that it possesses propagating ultra long lived modes.
}

\begin{document} 

\nolinenumbers

\maketitle

%%%%%%%%%%%%%%%%%%%%%%%%%%%%%%%%%%%%%%%%%%%%%%%%%%%%%%%%%%%%%%%%%%%%%%%%%%%%%%%
\section{Introduction}
%%%%%%%%%%%%%%%%%%%%%%%%%%%%%%%%%%%%%%%%%%%%%%%%%%%%%%%%%%%%%%%%%%%%%%%%%%%%%%%
%
In recent years the new age of multi-messenger gravitational wave astronomy has 
begun with the detection of gravitational waves from black hole mergers 
\cite{Abbott:2016blz, Abbott:2016nmj,Abbott:2017oio,Abbott:2017vtc,LIGOScientific:2018mvr}, 
and, in particular, the detection of a neutron star (NS) merger
\cite{Abbott:2017xzg}, whose electromagnetic signatures
have been observed as well
\cite{Coulter:2017wya,TheLIGOScientific:2017qsa,GBM:2017lvd,Abbott:2018wiz}. 
With this new channel of observation, 
it is possible to directly test the strong gravity regime. 
Therefore, the study of the properties of the gravitational waves 
emitted from astrophysical sources in 
well-motivated viable theories of gravity 
is of paramount importance.
	
Gravitational waves from merging compact objects are expected 
to exhibit a ringdown phase after the merger. 
During this phase, the gravitational perturbations oscillate 
with some characteristic frequencies while being exponentially 
attenuated
in time with some characteristic damping time. 
Theoretically the ringdown phase can be studied 
in terms of quasi-normal modes (QNMs) 
\cite{Kokkotas:1999bd,Nollert:1999ji,Berti:2009kk,Konoplya:2011qq}. 
Observationally, however, 
the precision of the detected signals is not yet sufficient
to extract the QNMs, although the next generation of instruments is 
expected to achieve the necessary sensitivity %
\cite{Berti:2018vdi,Barack:2018yly}.

The spectrum of QNMs of NSs has been studied 
in general relativity (GR) for a long time 
\cite{Andersson:1996pn,Andersson:1997rn,Kokkotas:1999mn,Benhar:2004xg,BlazquezSalcedo:2012pd,Blazquez-Salcedo:2013jka,Mena-Fernandez:2019irg,Volkel:2019gpq}. 
When the background spacetime is assumed to be static 
and spherically symmetric, the perturbations decouple into 
two independent channels, the axial (odd-parity) channel
and the polar (even-parity) one.
Axial perturbations lead to pure spacetime oscillations, 
and include the so called rapidly damped $w$-modes \cite{Kokkotas:2003mh}. 
Polar perturbations couple to the matter of the star, 
making their spectrum much richer. 
In GR there is no gravitational monopolar or dipolar radiation. 
Thus NSs typically have a ringdown spectrum dominated 
by the quadrupolar pressure-driven modes
consisting of the fundamental $f$-mode and the excited $p$-modes, 
that are supported by the fluid. 
They also feature polar spacetime modes, i.e., another set of $w$-modes. 

However, these are not all the modes the stars possess. 
It is well known that NSs in GR are characterized by a set of 
radial \textit{normal modes}, i.e., not damped modes, 
that become unstable modes beyond the maximum mass NS 
for a given equation of state (EOS) 
\cite{Chandrasekhar:1964zz,Chandrasekhar:1964zza,1966ApJ...145..505B,1966ApJ...145..514M,1977ApJ...217..799C,1983ApJS...53...93G,1992A&A...260..250V,1998IJMPD...7...49D}. 
{These modes consist of a fundamental $F$-mode, and the excited $H_N$-modes.}
Clearly in GR, radial perturbations cannot propagate gravitational radiation 
outside of the NS, and thus these modes play no role 
during the ringdown phase.

The purpose of this letter is to show that this scenario changes fundamentally
when dealing with NSs in alternative theories of gravity,
that introduce extra degrees of freedom 
in the form of additional fields, which enrich the spectrum of QNMs
even in the absence of matter 
(for a review see e.g.~\cite{Blazquez-Salcedo:2018pxo}).
As an interesting example, we here investigate NSs in
a class of scalar-tensor theories (STTs), the massive Brans-Dicke theory, where an ultra-light boson is non-minimally coupled with gravity \cite{Roshan:2011kz},
employing two potentials, a simple mass term and an exponential potential
(as obtained from $R^2$ gravity) 
\cite{Sotiriou:2008rp,Peebles:2002gy,Olmo:2005zr}.
NS configurations carry
a non-trivial massive scalar field, that significantly changes 
the properties of these stars 
\cite{Alsing:2011er,Astashenok:2013vza,Farinelli:2013pza,Orellana:2013gn,Staykov:2014mwa,Yazadjiev:2014cza,Yazadjiev:2015zia,Astashenok:2014pua,Astashenok:2014nua,Astashenok:2014dja,Astashenok:2017dpo,Sperhake:2017itk,Kase:2019dqc}.

In theories with gravitational scalar fields, 
axial QNMs of NSs are pure space-time modes 
(see \cite{Blazquez-Salcedo:2015ets,Blazquez-Salcedo:2018tyn,AltahaMotahar:2018djk,AltahaMotahar:2019ekm,Blazquez-Salcedo:2018qyy} 
for studies in various theories). 
Polar modes, in contrast, couple matter, scalar and tensor perturbations, 
making them  much more interesting and 
at the same time much more complicated to study.
Therefore the fundamental quadrupole mode
has so far only been obtained in the Cowling approximation
\cite{Sotani:2004rq,Yazadjiev:2017vpg,Staykov:2015cfa},
where all gravitational degrees of freedom are frozen.
{Full calculations have only been performed for 
the radial modes of scalarized NSs
\cite{Mendes:2018qwo,Doneva:2020csi}
showing the presence of scalar QNMs in massless STT and TMST,
and thus an enriched spectrum with respect to GR.}

{In this work we present the first calculation of the full QNM spectrum of NSs in scalar-tensor theory with ultra-light bosons, without making use of additional approximations. This allows us to explicitly show that the ringdown can be very different}
from the standard ringdown of GR and massless STTs:
when perturbing all the degrees of freedom in 
massive STT, 
the radial normal modes of GR 
turn into ultra long lived QNMs in STT.
%
%%%%%%%%%%%%%%%%%%%%%%%%%%%%%%%%%%%%%%%%%%%%%%%%%%%%%%%%%%%%%%%%%%%%%%%%%%%%%%%
\section{The theory}
%%%%%%%%%%%%%%%%%%%%%%%%%%%%%%%%%%%%%%%%%%%%%%%%%%%%%%%%%%%%%%%%%%%%%%%%%%%%%%%%
%
%
In massive STT the action in the Einstein frame with metric $g_{\mu\nu}$ and scalar field $\phi$ is given by \cite{Wagoner:1970vr}
\begin{equation}
S = \frac{1}{16\pi } \int d^4x \sqrt{-g} 
\big[ R - 2(\partial\phi)^2 - V(\phi) + L_{M}(A(\phi)^2 g,\chi) \big],
\label{EinsteinAction}
\end{equation}
where we choose the standard Brans-Dicke coupling function
$
A(\phi)= e^{-\frac{1}{\sqrt{3}}\phi} \
~.
$
We will consider two simple examples of potentials $V(\phi)$ \cite{Faraoni:2009km}: a simple mass term $V_I= 2 m_{\phi}^2 \phi^2$, and a potential of the form $V_{II}=\frac{3m_{\phi}^2}{2} \big(1- e^{-\frac{2\phi}{\sqrt{3}}}\big)^2$, which is related with $R^2$ gravity in the Jordan frame \cite{Faraoni:1999hp,Yazadjiev:2014cza,Yazadjiev:2015zia,Bhattacharya:2017pqc}. In practice we will see that both potentials give quantitatively very similar results.  
The corresponding field equations are
\begin{eqnarray}
G_{\mu\nu} = T^{(S)}_{\mu\nu} + 8 \pi T^{(M)}_{\mu\nu}
-V(\phi)g_{\mu\nu}/2
~, \nonumber \\
\nabla_{\mu}\nabla^{\mu}\phi = 
-4\pi\frac{1}{A}\frac{dA}{d\phi} T^{(M)} + \frac{1}{4} \frac{dV}{d\phi}
~,
\label{eq_G}
\end{eqnarray}
where $G_{\mu\nu} = R_{\mu\nu} - \frac{1}{2}Rg_{\mu\nu}$, 
$ T^{(S)}_{\mu\nu}=2\partial_{\mu}\phi\partial_{\nu}\phi -
g_{\mu\nu} \partial^{\sigma}\phi\partial_{\sigma}\phi $
and 
$ T^{(M)}_{\mu\nu} = (\rho + p)u_{\mu}u_{\nu} + pg_{\mu\nu} $ 
(i.e., the stress-energy tensor for a perfect fluid 
with energy density $\rho$, pressure $p$ and 4-velocity $u_{\mu}$).
The relation between the energy density and the pressure 
is determined by an EOS in the Jordan frame 
\cite{Faraoni:1999hp,Yazadjiev:2014cza,Yazadjiev:2015zia,Bhattacharya:2017pqc},
thus we have $\hat{\rho}=\hat{\rho}(\hat{p})$, with 
$ p = A^4 \hat{p}, \ \rho = A^4 \hat{\rho} $.

%%%%%%%%%%%%%%%%%%%%%%%%%%%%%
\section{The background}
%%%%%%%%%%%%%%%%%%%%%%%%%%%%%
%
For static and spherically symmetric NSs
we employ the standard Ansatz for the metric 
$
ds^2 = g_{\mu\nu}^{(0)} dx^{\mu} dx^{\nu} = -e^{2\nu(r)} dt^2 + e^{2\lambda(r)} dr^2 + r^2 d\Omega^2
$.
The scalar field, energy density and pressure are simply given by 
$\phi=\phi_0(r)$, $\hat{\rho}=\hat{\rho}_0(r)$ and $\hat{p}=\hat{p}_0(r)$, 
respectively, while the four-velocity of the static fluid 
is $u^{(0)}=-e^{\nu}dt$.
The NS configuration satisfies regularity 
at its center and at the border, defined by the point $r=R_s$ 
where the pressure vanishes. For asymptotically flat solutions the scalar field decays exponentially with $\phi \sim \frac{1}{r}e^{-m_{\phi} r}$, and the metric like $e^{2\nu}=e^{-2\lambda}\sim 1-2M/r$, 
where $M$ is the total mass of the NS.

%%%%%%%%%%%%%%%%%%%%%%%%%%%%%
\section{Polar perturbations} 
%%%%%%%%%%%%%%%%%%%%%%%%%%%%%
%
We assume a mode-expansion of the perturbation in terms 
of the spherical harmonics $Y_{lm}$ (with integer $l$, $m$) 
and the complex wave frequency $\omega$,
where $\omega = 2 \pi \omega_{R} + i/\tau$ 
with $\omega_R, \tau \in \mathbb{R}$, 
corresponding to the characteristic frequency 
and characteristic time of the mode, respectively. 
Hence the perturbations of each field can be written as
$\psi = \psi_{0} + \epsilon \sum \int d\omega e^{i\omega t}\delta \psi$, 
with the fields $\psi=\{g,u,\phi,\hat \rho,\hat p\}$ being the metric, 
4-velocity, scalar field, energy density and pressure, respectively, 
perturbed up to first order in $\epsilon$. 
The sum is over all $l$ and $m$.
As mentioned, the perturbation can be separated into axial and polar parts \cite{Kokkotas:1999bd,Nollert:1999ji,Berti:2009kk,Konoplya:2011qq,FernandezJambrina:2003mv}.

Focusing on the polar channel, the metric perturbation is given by
$
\delta g = 
\big[
r^l e^{2\nu} H_0 Y_{lm} dt^2 -2 i \omega r^{l+1} H_1 Y_{lm} dt dr + r^l e^{2\lambda} H_2 Y_{lm} dr^2 + r^{l+2} K  Y_{lm} d\Omega^2
\big]
$; the 4-velocity perturbation is
$
\delta u = \frac{1}{2} r^l e^{\nu} H_0 Y_{lm}  dt + r^l i\omega e^{-\nu} \left(e^{\lambda}W/r -r H_1 \right) Y_{lm} dr -i\omega r^l e^{-\nu} V_1 (\partial_{\theta} Y_{lm} d\theta+\partial_{\varphi} Y_{lm} d\varphi)
$; for the scalar field, energy density and pressure perturbations, 
we have $\delta \phi = \phi_1 Y_{lm}$, $\delta \hat \rho = E_1 Y_{lm}$ 
and $\delta \hat p = \Pi_1 Y_{lm}$, respectively. 

Introducing this Ansatz for the perturbations in the field equations 
(\ref{eq_G}), one obtains a system of  
ordinary differential equations (ODEs) in $r$, 
characterized by the eigenvalue $\omega$ and the multipole number $l$.
After some tedious algebraic manipulations and introducing the function,
$
X={\omega}^{2} \left(\hat{p}_0 +\hat{\rho}_0  
\right) {{e}^{-\nu}}{V_1} 
-\frac{1}{r}{\frac{d\hat{p}_0}{{d}r}} {{e}^{\nu-\lambda}}{W}+\frac{1}{2}\left( \hat{p}_0+\hat{\rho}_0\right) {{e}^{\nu}}{H_0} 
$ \cite{Lindblom:1983ps},
one can show that the minimal system of ODEs is given by a set 
of six first order ODEs for the functions 
$\Psi = \{K,  H_1, W, X, \phi_1, \frac{d\phi_1}{dr}\}$, 
taking the form
\begin{equation}
\label{eq_Psi}
\frac{d}{dr} \Psi + \sigma\Psi = 0~,
\end{equation}
where $\sigma$ is a matrix that depends in a complicated way 
on the static functions $\{\nu, \lambda, \phi_0, \hat{p}_0, \hat{\rho}_0\}$, 
the eigenvalue $\omega$ and multipole number $l$.

Note that inside the star, the perturbation is described by the functions $\{K, H_1\}$ (metric),  
$\{W, X\}$ (fluid), 
and $\{\phi_1,\frac{d\phi_1}{dr}\}$ (scalar). 
Outside the star, since there is no fluid $\hat{p}=\hat{\rho}=0$, 
also $W=X=0$, and the system reduces 
to four first order ODEs
for metric and scalar perturbations.

Note that in STT all perturbation equations 
are coupled with each other. 
In the GR limit, formally obtained when $m_{\phi}\to\infty$, 
the scalar field vanishes and the system of ODEs 
decouples into two parts, 
one for the metric and fluid perturbations, 
and the second for the scalar field perturbation (a minimally coupled 
scalar test field equation of GR).

%%%%%%%%%%%%%%%%%%%%%%%%%%%%%
\section{Asymptotic behaviour of the gravitational wave}
%%%%%%%%%%%%%%%%%%%%%%%%%%%%%
%
%
At infinity, the massive scalar field is suppressed exponentially. 
Therefore the scalar perturbation is effectively asymptotically 
decoupled from the metric perturbations. 
Sufficiently far from the star, 
the perturbation can thus be described in terms of two decoupled 
Schr\"odinger-like equations 
$ \frac{d^2Z_{g,s}}{dy^2}= (G_{g,s}(r)-\omega^2) Z_{g,s} $, 
with tortoise coordinate $\frac{dy}{dr}=e^{\lambda-\nu}$:
one for gravitational perturbations (given by the potential $ G_g(r)=
\frac{2(r-2M)}{r^4(nr+3M)^2}\left(n^2(n+1)r^3+3Mn^2r^2+9M^2nr+9M^3\right) $, 
with $n=l(l+1)/2$, and Zerilli function $Z_g(r)$, 
which is a linear combination of the functions 
$H_1$ and $K$ \cite{Zerilli:1970se}), 
and the other one for scalar perturbations (with the potential 
$G_s(r)=(1-\frac{2M}{r})\left(\frac{2}{r^2}
(n+\frac{M}{r})+m_{\phi}^2\right)$
and function $ Z_s(r)=\phi_1(r)/r^l$).

For space-time perturbations the potential goes to zero at infinity,
$G_g(r\to\infty)\to 0$, implying that waves traveling away 
from the star behave like $Z_g \sim 
e^{i\omega y}$.
However, for scalar perturbations, the potential goes like 
$G_s(r\to\infty)\to m_{\phi}^2$, implying that outgoing scalar waves 
behave like $Z_s \sim 
e^{i\Omega y}$,
with $\Omega$ given by %
$ \Omega^2 = \omega^2 - m_{\phi}^2 $.

To calculate the QNMs, we first obtain numerically the background solution 
solving the static equations %
on a compactified coordinate $x=\frac{r}{R_s+r}$ \cite{Ascher:1979iha}.
Then the perturbation equation (\ref{eq_Psi}) 
is integrated for two independent solutions 
from the center of the star up to some point $r_i>R_s$. 
We choose this $r_i$ so that $\phi_0(r_i)\lesssim 10^{-4}$. 
The QNMs are determined by matching 
at $r=r_i$ these two solutions 
of eq.~(\ref{eq_Psi}) with outgoing wave solutions 
of the Zerilli and scalar perturbations for the Schwarzschild background.

\begin{figure}
	\onefigure[width=1\linewidth,angle =-90]{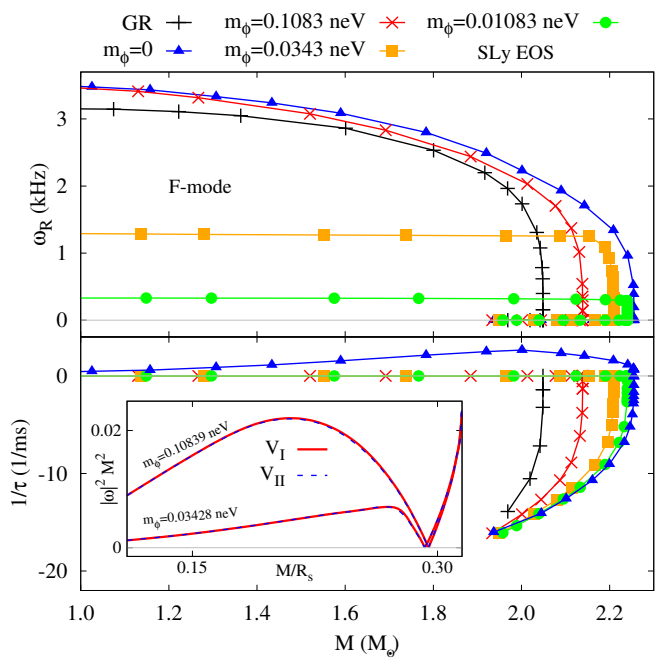}
	\caption{{The $F$-mode}:
		$\omega_R$ in kHz (top) and $1/\tau$ 
		in 1/ms (bottom) vs $M$ in $M_{\odot}$ 
		for several scalar masses $m_\phi$, and SLy EOS. 
		For comparison the GR and massless STT modes are shown. 
		Inset: $|\omega|^2 M^2$ vs 
		compactness $M/R_s$ for the potentials $V_I$ (solid red line) 
		and $V_{II}$ (dashed blue line) and two values of the scalar mass.
	}
	\label{fig:radial_p-modes_sly}
\end{figure}
 
%%%%%%%%%%%%%%%%%%%%%%%%%%%%%%%%%%%%%%%\textit%%%%%%%%%%%%%%%%%%%%%%%%%%%%%%%%%
 \section{Ultra long lived modes}
%%%%%%%%%%%%%%%%%%%%%%%%%%%%%%%%%%%%%%%%%%%%%%%%%%%%%%%%%%%%%%%%%%%%%%%%%%%%%% 
%
The presence of the scalar degree of freedom in massive STT
leads to a much richer spectrum of modes as compared to GR.
Let us first focus on $l=0$ modes,
that contribute to the gravitational radiation of the star 
in STT.
Essentially there are two families of modes, 
that we name according to their GR limits.
{First, we have modes related to oscillations of the NS matter: the fundamental pressure-led mode ($F$ mode) and its excitations ($H_1$, $H_2$, etc). 
Second, we have modes related to oscillations of a
minimally coupled scalar field in GR: the scalar-led $\phi$-modes.}

In GR, 
{these two families of modes} are
completely decoupled.
They form the spectrum of two different perturbation equations.
{The pressure-led modes}
represent \textit{normal modes} in GR, since
these radial fluctuations cannot propagate outside the star.
{The $F$-mode} becomes a zero mode 
for the maximum mass neutron star,
and an unstable mode beyond the maximum mass.
When a scalar field is minimally coupled to GR,
{$\phi$-modes} arise, that do propagate outside the star.
These {$\phi$-modes} are damped QNMs.

Since in STTs the scalar field is a gravitational degree of freedom,
intimately coupled with the tensor degrees of freedom,
the GR pressure-led modes turn into modes
that are no longer purely normal {(in the spirit of the toy model studied in \cite{Kokkotas:1986gd})}. Indeed, the scalar field
allows for gravitational $l=0$ radiation. 
In the presence of a gravitational scalar field the 
 pressure-led modes
then turn into damped QNMs.
This is demonstrated in Fig.~\ref{fig:radial_p-modes_sly}
for the 
{$F$-mode} in GR (black) and in a STT 
with a massless scalar field (blue), for the SLy EOS.

\begin{figure}
\onefigure[width=0.85\linewidth,angle =-90]{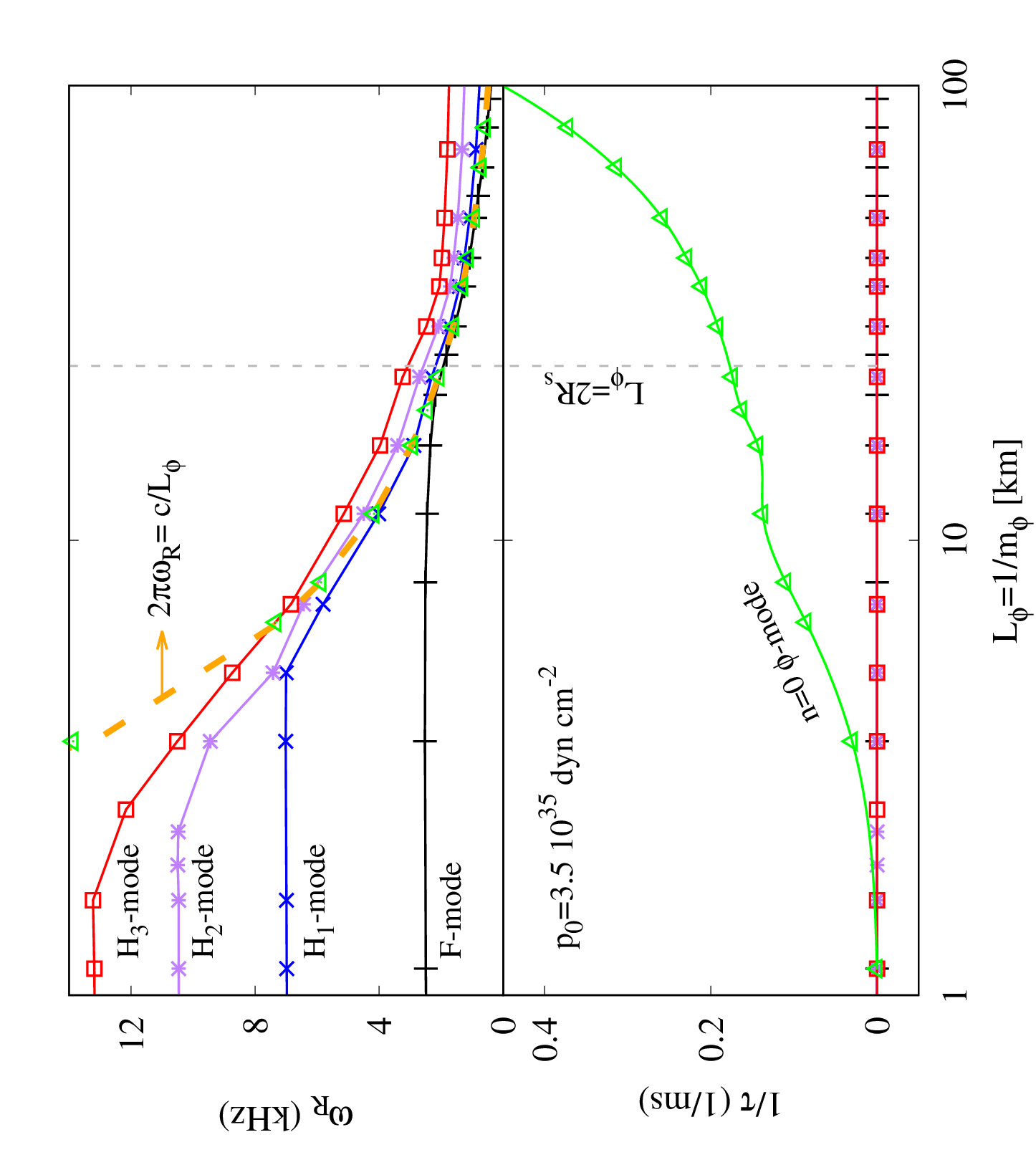}
\caption{{The
	 $F$, 
	 $H_1$, 
	 $H_2$ and
	 $H_3$ modes} (in black, blue, purple and red respectively) and the fundamental $l=0$ $\phi$-mode (green):
$\omega_R$ in kHz (top) and $1/\tau$
in 1/ms (bottom) vs Compton wavelength $L_\phi=1/m_\phi$ in km 
for fixed central pressure $p_0$ (SLy EOS) and
potential $V_I$. 
For comparison (top) the inverse of the 
Compton wavelength $L_\phi$ (orange) and $L_\phi=2R_s$ (grey) is shown.
Potentials $V_I$ and $V_{II}$ 
give very similar results.
}
\label{fig:ground_mode_sly}
\end{figure}

{The $F$-mode} is also shown in Fig.~\ref{fig:radial_p-modes_sly}
for massive STT for several values of the scalar mass $m_\phi$
in the range of ultra light scalars.
The here considered masses reside in
the allowed range of the free parameter of the theory.
Note that the observational bound \cite{Naf:2010zy,Brito:2017zvb}
for STT with ultralight bosons requires typically $m_\phi \gtrsim 10^{-13}$ eV.
For large scalar mass $m_\phi$, GR is approached.
In fact, for $m_\phi \gtrsim 10^{-9}$ eV little deviation
from GR is observed. In the inset we demonstrate that modes of potential $V_I$ (solid red line) and $V_{II}$ (dashed blue line) are almost indistinguishable. 

In this range
of scalar masses the frequency $\omega_R$
of the {$F$-mode}
undergoes a fundamental change: for larger $m_\phi$ the 
dependence on the NS mass is similar to GR, 
except that higher maximal NS masses are reached, 
whereas for smaller $m_\phi$ the frequency becomes rather
independent of the NS mass, except in the region close to the
maximal NS mass. 
This is illustrated in Fig.~\ref{fig:ground_mode_sly},
where for a family of stars with fixed central pressure
the Compton wavelength $L_\phi$ is varied.
In the Figure we show the {$F$-mode} (black), three excited states {($H_1$ in blue, $H_2$ in purple and $H_3$ in red)}, and the fundamental {$\phi$-mode} (green). We also show the frequency given by the inverse of the Compton wavelength $2\pi \omega_R = c/L_{\phi}$.

For small Compton wavelength $L_\phi$ 
the scale for $\omega_R$ of the (excited) pressure-led modes is set by (multiples of) the size of the
star, and the mode decays exponentially with distance in a range of $10-100$ km.

The {$F$-mode} undergoes a change precisely when the Compton wavelength
$L_\phi=1/m_\phi$ of the scalar field matches the size of the star, 
$L_\phi=2R_s$ (i.e. $m_{\phi}=0.052$ neV). In contrast, for larger $L_\phi$ 
the scale is set by the mass of the scalar field (orange),
in fact, $2 \pi \omega_R \approx m_\phi$.

For the {$H_N$-modes}, this change in behaviour occurs at lower values of the Compton wavelength (i.e. $m_{\phi}=0.072, 0.16, 0.19$ neV for the first three excitations). 
In the Figure we can see that for large values of the Compton wavelength, $2 \pi \omega_R > m_\phi$. 
Our numerical estimations for the propagating distances of these modes indicate that they should be equal or larger than $\sim 10^5$ ly, meaning these modes could propagate at least within our galaxy, if not further.

As noticed previously for the background \cite{Yazadjiev:2014cza},
the scalar field increases in significance with 
increasing Compton wavelength $L_\phi$.
Here we see that this also holds for the pressure perturbations.
For small $L_\phi$ the pressure perturbations
are almost decoupled from the scalar perturbations,
and the {$F$-mode and $H_N$-modes} are very close to GR.
Thus for the pressure modes the pressure perturbation functions 
are larger than the scalar ones.
This changes when the Compton wavelength $L_\phi$
exceeds the size of the star.
Then the equations are strongly coupled,
and the amplitudes of the pressure and scalar perturbation
functions are of the same order.

The imaginary part of the {$F$-mode and the $H_N$-modes}, $\omega_I$, 
on the other hand, remains extremely small within the allowed range
of $m_\phi$. Our calculations indicate that the decay time
$\tau$ is larger than or equal to $\sim 10^5$ y. 
Thus these novel modes are ultra long lived.
In contrast to long lived modes of black holes
\cite{Konoplya:2004wg}
the modes continue to exist all the way up to the limit $m_\phi \to \infty$,
where they become normal modes of the GR stars.
We note that long lived scalar radiation has also been seen during 
the core collapse process in massive STTs \cite{Sperhake:2017itk}.

For comparison,
the fundamental {$\phi$-mode} is also shown in Fig.~\ref{fig:ground_mode_sly}.
Its frequency is always set by the Compton wavelength $L_\phi$.
Thus the {$F$-mode and fundamental $\phi$-mode} frequencies are very close
for $L_\phi \gtrsim 2 R_s$. 
The decay time $\tau$ is, however, very different for both
types of modes. In contrast to the ultra long lived {$F$-mode and $H_N$-modes,
the $\phi$-modes} have a decay time on the order of ms, and they decay with distance in $10-100$ km.

% 
%
%

%%%%%%%%%%%%%%%%%%%%%%%%%%%%%%%%%%%%%%%%%%%%%%%%%%%%%%%%%%%%%%%%%%%%%%%%%%%%%%%
\section{The quadrupolar fundamental mode}
%%%%%%%%%%%%%%%%%%%%%%%%%%%%%%%%%%%%%%%%%%%%%%%%%%%%%%%%%%%%%%%%%%%%%%%%%%%%%%%%
%
In GR the $l=2$ fundamental mode may be the most interesting mode 
with regard to astrophysical scenarios, since simulations show 
that it tends to dominate the ringdown spectrum after a merger. 
Previously, analysis of this mode has been performed in STT
\cite{Sotani:2004rq,Sotani:2005qx} and $R^2$ gravity \cite{Staykov:2015cfa}
only in the Cowling approximation.

\begin{figure}
\onefigure[width=0.85\linewidth,angle =-90]{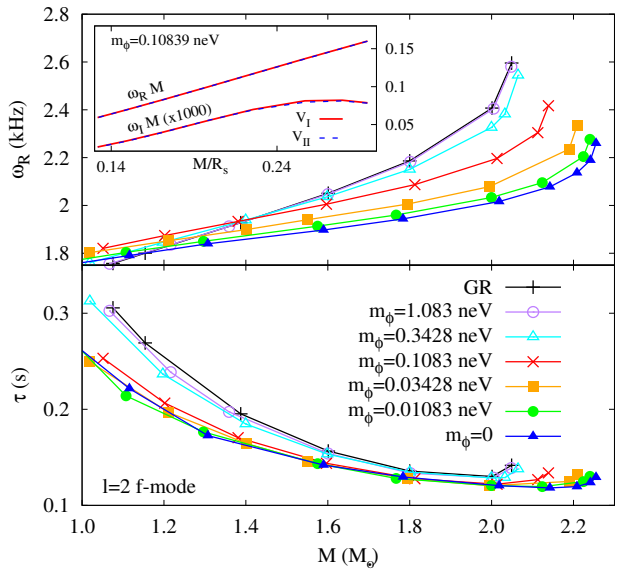}
\caption{The $l=2$ fundamental mode:
$\omega_R$ in kHz (top) and $\tau$
in s (bottom) vs $M$ in $M_{\odot}$
for several scalar masses $m_\phi$ and SLy EOS.
For comparison the GR and massless STT modes are shown.
Inset: comparison between $V_I$ (solid red line)
and $V_{II}$ (dashed blue line).
}
\label{fig:fmode_sly}
\end{figure}

In Figure \ref{fig:fmode_sly} we show 
the frequency $\omega_R$ (top) and the damping time $\tau$ (bottom)
for the $l=2$ fundamental mode, i.e., the $f$-mode,
for several values of the scalar mass $m_{\phi}$, using the SLy EOS. 
Also shown are the corresponding results for GR (black)
and a STT with $m_\phi=0$ (blue).
The chosen mass values are again in the physically interesting range,
$m_{\phi}=1.08$, $0.343$, $0.108$, $0.0343$ and $0.0108$ neV,
compatible with current observations and constraints \cite{Brito:2017zvb}.
Note that for $1.08$ neV the values are very close to GR already, 
and values below $m_{\phi}=1$ peV do not deviate much from the 
$m_{\phi}=0$ ones.

In general, the behaviour is reminiscent of previous observations for axial modes \cite{Blazquez-Salcedo:2018qyy}, with the frequency and the damping time, typically varying within $10\%$ of the GR value. Fig.~\ref{fig:fmode_sly} shows that a decrease of the scalar mass $m_\phi$ 
leads to a decrease of the frequency, except for low values of the NS mass $M$ (close to $1 M_\odot$),
where the frequency rises slightly above the GR value. 
A decrease of $m_\phi$ leads also to a decrease 
of the damping time $\tau$.

Comparing the results from the Cowling approximation \cite{Staykov:2015cfa} 
with the full calculations, we note, that the dependence 
of the frequency on the scalar mass $m_\phi$ is much smoother now.
In the Cowling approximation, the frequency for
the larger values of $m_\phi$ first increases with decreasing
$m_\phi$, and decreases (mostly) only for the smaller values of 
$m_\phi$ in the considered mass range.
This shows the relevance of performing the full calculations
also for the $f$-mode, apart from obtaining the decay time, of course.

%%%%%%%%%%%%%%%%%%%%%%%%%%%%%%%%%%%%%%%%%%%%%%%%%%%%%%%%%%%%%%%%%%%%%%%%%%%%%%%
\section{Conclusions}
%%%%%%%%%%%%%%%%%%%%%%%%%%%%%%%%%%%%%%%%%%%%%%%%%%%%%%%%%%%%%%%%%%%%%%%%%%%%%%%%% 
The spectrum of gravitational waves emitted from NSs in massive STT
is much richer than the corresponding GR spectrum,
since the theory effectively introduces 
an additional scalar degree of freedom.
Consequently, the quadrupole modes of GR
are modified, turning into gravitational-led modes.
These are augmented by a new class of gravitational modes, the scalar-led
quadrupole modes. 

Moreover, {we have shown that a novel type of} 
monopolar gravitational radiation arises, 
not present in GR {nor in massless STTs}. 
The radial normal modes of the NS matter
in GR turn into propagating QNMs in STT. 
For scalar masses beyond $\sim 0.1$ neV, 
we have shown that the modes are not trapped, 
and lead to gravitational wave radiation. 
Intriguingly, all these modes are ultra long lived
and the frequencies pass precisely
the LIGO/VIRGO sensitivity window.
{It is important though, that a mode also has a degree of excitation
that will lead to an observable signal. We will therefore address
this highly relevant question in the future.}
The pressure-led modes are again augmented by 
scalar-led modes, which, in contrast, possess damping times
in the range of ms.

The next steps will be to further complete the spectrum of QNMs
in STT, and to address NS universal relations for the polar modes. 
A similar effect is expected to appear with spontaneous scalarization 
type of couplings. Then effects of rotation should be included,
which may mix the monopolar and quadrupolar modes,
due to the reduction of symmetry.
Numerical simulations of mergers in GR show 
that the gravitational wave emission after the merger is described 
as a ringdown dominated by three modes 
\cite{Bauswein:2015yca,Bernuzzi:2015rla}. 
Two of these modes are usually interpreted as the fundamental mode 
and first excitation of a mixture of the radial modes of the star 
and the fundamental quadrupolar mode \cite{Stergioulas:2011gd,Takami:2014tva,Bauswein:2019qgh,Vretinaris:2019spn}. 
While merger simulations have mainly been done in GR, 
recently an effective model was used to estimate the merger 
of NSs in $R^2$ gravity \cite{Sagunski:2017nzb}, 
observing two modes that dominate the ringdown.
It will be interesting to analyze and interpret
the dominant frequencies in full models of mergers in massive STTs, 
and their dependence on the mass of the scalar field.

%%%%%%%%%%%%%%%%%%%%%%%%%%%%%%%%%%%%%%%%%%%%%%%%%%%%%%%%%%%%%%%%%%%
\acknowledgments
%%%%%%%%%%%%%%%%%%%%%%%%%%%%%%%%%%%%%%%%%%%%%%%%%%%%%%%%%%%%%%%%%%%
The authors gratefully acknowledge support by the DFG funded
Research Training Group 1620 ``Models of Gravity'' 
and the COST Actions CA15117 and CA16104. %
JLBS would like to acknowledge support from the DFG project BL 1553.
FSK acknowledges previous support by the Croatian Science Foundation 
under the Project IP-2014-09-3258, and the European Union 
through the European Regional Development Fund - the Competitiveness 
and Cohesion Operational Programme (KK.01.1.1.06).
FSK thanks the hospitality of University of Oldenburg 
during part of this project,
where she was also partially supported by the RBI institutional financing for the purpose of short-term visits.

%%%%%%%%%%%%%%%%%%%%%%%%%%%%%%%%%%%%%%%%%%%%%%%%%%%%%%%%%%%%%%%%%%%
%
%

%
\bibliography{STT_bib_c}

\end{document}